\documentclass[aps,prl,reprint,groupedaddress]{revtex4-2}

\usepackage{natbib}
\usepackage{graphicx}   
\usepackage{xcolor}
\usepackage{soul}       
\usepackage{orcidlink}

\begin{document}

\newcommand{\pdcz}{pulse-driven convective zone}

\newcommand{\spr}{\mbox{$s$-process}}
\newcommand{\sprn}{\mbox{$s$ process}}
\newcommand{\ipr}{\mbox{$i$-process}}
\newcommand{\iprn}{\mbox{$i$ process}}
\newcommand{\npr}{\mbox{$n$-process}}
\newcommand{\nprn}{\mbox{$n$ process}}
\newcommand{\rpr}{\mbox{$r$-process}}
\newcommand{\rprn}{\mbox{$r$ process}}
\newcommand{\ppr}{\mbox{$p$-process}}
\newcommand{\pprn}{\mbox{$p$ process}}


%

\newcommand{\unitspace}{\ensuremath{\,}}
\newcommand{\usp}{\unitspace}
\newcommand{\numberspace}{\ensuremath{\;}}
\newcommand{\nsp}{\numberspace}
\newcommand{\unitstyle}[1]{\ensuremath{\mathrm{#1}}}
\newcommand{\power}[2]{\ensuremath{{#1}^{#2}}}
\newcommand{\natlog}[2]{\ensuremath{#1\times 10^{#2}}} 
\newcommand{\ee}[1]{\ensuremath{\times 10^{#1}}}

\newcommand{\nano}{\unitstyle{n}}
\newcommand{\milli}{\unitstyle{m}}
\newcommand{\centi}{\unitstyle{c}}
\newcommand{\kilo}{\unitstyle{k}}
\newcommand{\Mega}{\unitstyle{M}}
\newcommand{\Giga}{\unitstyle{G}}

\newcommand{\meter}{\unitstyle{m}}
\newcommand{\kilogram}{\kilo\gram}
\newcommand{\second}{\unitstyle{s}}

\newcommand{\Kelvin}{\unitstyle{K}}
\newcommand{\K}{\Kelvin}  

\newcommand{\cm}{\centi\meter}
\newcommand{\gram}{\unitstyle{g}}

\newcommand{\grampercc}{\gram\usp\power{\cm}{-3}} 
\newcommand{\grampersquarecm}{\gram\usp\power{\cm}{-2}} 
\newcommand{\squarecmpergram}{\power{\cm}{2}\usp\power{\gram}{-1}} 
\newcommand{\GramPerCc}{\grampercc}
\newcommand{\GramPerSc}{\grampersquarecm}
\newcommand{\columnunit}{\grampersquarecm}
\newcommand{\dyne}{\unitstyle{dyn}} 
\newcommand{\erg}{\unitstyle{ergs}} 
\newcommand{\ergs}{\erg}
\newcommand{\gauss}{\unitstyle{G}} 
\newcommand{\ergspersecond}{\erg\unitspace\power{\second}{-1}}
\newcommand{\ergspergram}{\erg\unitspace\power{\gram}{-1}}
\newcommand{\ergspergs}{\erg\unitspace\power{\gram}{-1}\unitspace\power{\second}{-1}} 
\newcommand{\ergssecond}{\erg\unitspace\second}
\newcommand{\cgsflux}{\erg\unitspace\power{\cm}{-2}\usp\power{\second}{-1}}

\newcommand{\amu}{\unitstyle{u}} 
\newcommand{\angstrom}{\mbox{\AA}} 
\newcommand{\fermi}{\unitstyle{fm}} 
\newcommand{\eV}{\unitstyle{eV}}        
\newcommand{\keV}{\kilo\eV} 
\newcommand{\MeV}{\Mega\eV} 

\newcommand{\Msun}{\ensuremath{\unitstyle{M}_\odot}}
\newcommand{\Lsun}{\ensuremath{\unitstyle{L}_{\odot}}}
\newcommand{\Rsun}{\ensuremath{\unitstyle{R}_{\odot}}}
\newcommand{\Zsun}{\ensuremath{Z_{\odot}}}
\newcommand{\Myr}{\Mega\yr}
\newcommand{\Gyr}{\Giga\yr}
\newcommand{\parsec}{\unitstyle{pc}}
\newcommand{\kpc}{\kilo\parsec} 
\newcommand{\mJy}{\unitstyle{\mu Jy}} 
\newcommand{\Msunyr}{\Msun\,\power{\yr}{-1}}
\newcommand{\MJ}{\ensuremath{\mathrm{M_J}}}
\newcommand{\RJ}{\ensuremath{\mathrm{R_J}}}
\newcommand{\AU}{\unitstyle{AU}}

\newcommand{\minute}{\unitstyle{min}} 
\newcommand{\hour}{\unitstyle{hr}} 
\newcommand{\yr}{\unitstyle{yr}}        
\newcommand{\km}{\kilo\meter}   
\newcommand{\Hz}{\unitstyle{Hz}}        
\newcommand{\ksec}{\kilo\second} 
\newcommand{\kms}{\ensuremath{\mathrm{km}\,\second^{-1}}\xspace}
\newcommand{\vcrit}{{\varv_{\mathrm{crit}}}}
\newcommand{\vkep}{{\varv_{\mathrm{kep}}}}
\newcommand{\vsurf}{{\varv_{\mathrm{surf}}}}
\newcommand{\mol}{\unitstyle{mol}}
\newcommand{\barn}{\unitstyle{b}} 

\newcommand{\unit}[2]{\ensuremath{#1\numberspace\mathrm{#2}}}


\newcommand{\fhcom}[1]{{\color{red}[\emph{FH: #1}]}}
\newcommand{\pdtxt}[1]{{\color{blue}[PD: #1]}}
\newcommand{\ebtxt}[1]{{\color{violet}[EB: #1]}}
\newcommand{\ebcom}[1]{{\color{violet}[\emph{EB: #1}]}}

\newcommand{\pasp}{\mbox{Publ. Astron. Soc. Pacific}}
\newcommand{\apjl}{\mbox{Astrophys. J. Lett.}}
\newcommand{\aj}{\mbox{Astron. J.}}
\newcommand{\araa}{\mbox{Annu. Rev. Astron. Astrophys.}}
\newcommand{\aap}{\mbox{Astron. Astrophys.}}
\newcommand{\aaps}{\mbox{Astron. Astrophys. Suppl.}}
\newcommand{\mnras}{\mbox{Mon. Not. R. Astron. Soc.}}


\title{Can COSI detect $\gamma$-ray lines from rare isotopes produced in the astrophysical intermediate neutron-capture process?}


\author{Falk Herwig\,\orcidlink{0000-0001-8087-9278}}
\author{Pavel Denissenkov\,\orcidlink{0000-0001-6120-3264}}
\affiliation{Astronomy Research Centre and Department of Physics and Astronomy}
\affiliation{University of Victoria, Victoria, British Columbia V8W 2Y2, Canada }
\affiliation{CaNPAN (Canadian Nuclear Physics for Astrophysics Network) Collaboration}
\affiliation{NuGrid Collaboration}

\author{Eric Burns\,\orcidlink{0000-0002-2942-3379}}
\affiliation{Department of Physics and Astronomy, Louisiana State University, Baton Rouge, LA 70803, USA}


\date{\today}

\begin{abstract}
We investigate the nuclear $\gamma$-ray line emission from rare
isotopes produced in the astrophysical intermediate neutron-capture
process (\iprn) and assess the prospects of observing these emissions
with $\gamma$-ray telescopes. The astrophysical sites of the
\iprn\ remain uncertain, but two candidates with predicted rapid mass
ejections at metallicities of stars in the solar neighborhood are
post-asymptotic giant branch (post-AGB) stars, such as Sakurai's
object (V4334 Sagittarii), and rapidly-accreting white dwarfs
(RAWDs). Detailed 1D and 3D simulations of these scenarios indicate
that the convective-reactive astrophysical fluid dynamics responsible
for \ipr\ nucleosynthesis can lead to violent, non-radial outbursts
that ultimately result in mass ejections of \ipr\ products. We
calculate the ejected yields of rare isotopes whose radioactive decays
may produce detectable $\gamma$-ray lines, particularly in the 0.5–2
MeV energy range. Our analysis focuses on isotopes such as $^{22}$Na,
$^{89}$Sr, and $^{95}$Zr, which are expected to generate long-lasting
emissions potentially observable by the COSI $\gamma$-ray
telescope. We estimate the formation rates of these sources and the
likelihood of detecting their $\gamma$-ray emissions within 1000
parsecs of the Sun. We find that the probability of observing
\ipr\ emission lines during COSI's operational period is up to
$\approx 1\%$, but could rise to $11\%$ for $^{89}$Sr if the event is
observed within a few days. Due to the long lifetime and large
production of $^{22}$Na from proton-capture reactions its detection is
more likely, with a probability of $\approx 5\%$. Future space
missions could significantly enhance detection capabilities,
potentially increasing the observation probability to several tens of
percent. Detection of long-lived neutron-rich isotopes such as
  $^{137}$Cs would provide the first direct $\gamma$-ray signature of
  intermediate neutron-density nucleosynthesis, distinguishing the i
  process from classical s- and r-process pathways. These findings
outline a multi-messenger approach to studying dynamic stellar
neutron-capture nucleosynthesis through $\gamma$-ray observations.
\end{abstract}


\maketitle


\section{Introduction}
Multi-messenger astronomy is rapidly emerging as a powerful tool for
investigating time-domain astrophysics and reactive flow processes
\cite{burns2025}.  The new NASA $\gamma$-ray Compton Spectrometer and
Imager (COSI) space telescope, scheduled for launch in 2027, will have
a narrow-line sensitivity limit of approximately $3\times
10^{-6}\ \mathrm{photons\,cm}^{-2}\mathrm{s}^{-1}$ for point sources
at the $3\sigma$ level over a two-year survey, with energy coverage
spanning approximately $0.2$--$5\,\mathrm{MeV}$ \cite{tomsick2024}.
We focus on the $0.5$--$2\,\mathrm{MeV}$ band, where its narrow-line
sensitivity is best. Outside this band the sensitivity decreases due
to increased background and reduced effective area. COSI surveys the
full sky approximately every day ($\approx 13\,\mathrm{h}$), so any
sufficiently bright $\gamma$-ray line transient lasting longer than
$\sim 1\,$day will be observed without dedicated pointing
\cite{tomsick2024}.

COSI is expected to detect $\gamma$-ray lines from $^{56}$Co,
$^{44}$Ti, and $^{60}$Fe in known Type Ia and core-collapse supernova
remnants, improve upon COMPTEL and INTEGRAL/SPI measurements of the
1.809 MeV emission of $^{26}$Al from massive stars in the Milky Way
disk \cite{schonfelder2000,bouchet2015,pleintinger2023}, observe the
0.511 MeV positron annihilation line from the Galactic bulge and disk
\cite{siegert2016}, and search for lines from $^7$Be and $^{22}$Na
decays in nova explosions
\cite{siegert2018,fougeres2023natcomms22na}, provided such an event
occurs during its operational period.

We propose additional targets of opportunity for COSI observations:
$\gamma$-ray lines from the decays of relatively long-lived unstable
isotopes produced in the intermediate neutron capture process (\iprn) in post-asymptotic giant branch (post-AGB) stars and rapidly-accreting
white dwarfs (RAWDs).

The \iprn\ is a neutron-capture nucleosynthesis regime intermediate in
neutron density compared to the much slower \sprn\ \cite{busso1999}
and the much more rapid \rprn\ \cite{cowan2021}. It occurs in a star
when shell He ignition under partially degenerate conditions leads to
a thermo-nuclear runaway. If convective mixing induced by the thermal
flash reaches the boundary between the He-rich and H-rich zones,
proton ingestion can trigger the \iprn. The ingested protons initiate
the reaction chain
$^{12}$C(p,$\gamma)^{13}$N(e$^+\nu)^{13}$C($\alpha$,n)$^{16}$O,
producing neutron densities of $N_\mathrm{n}\sim
10^{13}$--$10^{15}\ \mathrm{cm}^{-3}$, depending on the H-mass
ingestion rate and the peak He-burning temperature
\cite{cowan1977,malaney1986}. These neutron densities are intermediate
between those characteristic of the $s$- and \rpr\ nucleosynthesis
\cite{kaeppeler2011,thielemann2011}. A recent review of the \iprn, including its stellar simulation context,
observational manifestations and the associated nuclear physics, is
given by \cite{Wiedeking2025}.

Direct evidence of the \iprn\ occurring in stars comes from the
enhancement of the first-peak elements Rb, Sr, Y, and Zr by
approximately two orders of magnitude in the post-AGB star Sakurai's
object (V4334 Sagittarii) \cite{asplund1999,evans2020sakurai}, where
the \iprn\ was triggered by a very late thermal pulse
\cite[VLTP,][]{herwig2001b} of its He shell
\cite{duerbeck2000,herwig2011}, and from the enhancement of trans-Fe
elemental abundances in H-deficient hot white dwarfs that have
undergone a He-core flash after leaving the red-giant branch (RGB) due
to extensive mass loss \cite{battich2023}.

There is also ample indirect evidence. The heavy-element abundance
patterns of most C-enhanced metal-poor stars classified as
CEMP-$r$/$s$ \cite{beers2005} are difficult to explain as simple
mixtures of $s$- and \rpr\ products \cite{bisterzo2012}, but are well
reproduced by
\ipr\ models, both simplified one-zone models with constant neutron
density \cite{dardelet2014,hampel2016,hampel2019} and realistic
multi-zone models of low-mass AGB stars \cite{choplin2022a} and RAWDs
\cite{denissenkov2019,stephens2021}. Anomalous isotopic ratios in
certain presolar graphite grains \cite{jadhav2013} and in SiC grains
of type AB \cite{fujiya2013} and mainstream type \cite{liu2014} may
also have originated from \ipr\ nucleosynthesis. An \ipr\ signature in
the Ba and La ratio of open clusters \cite{mishenina2015} suggests
that the \iprn\ can operate at high metallicities with neutron
exposures reaching the second peak, although the mechanism remains
unclear. The radioactive isotope $^{208}$Tl can also be produced in
the \iprn\ at neutron densities of $\sim
10^{15}\ \mathrm{cm}^{-3}$ \cite{vassh2024}, and its 2.6 MeV
$\gamma$-line emission could be detectable if actinides are
synthesized alongside it \cite{choplin2022b}.

In the following we use multi-zone \ipr\ models of a solar-metallicity
post-AGB VLTP, calibrated to conditions inferred for Sakurai's object
\cite{herwig2011} but used here as a generic VLTP scenario, and of
RAWDs to calculate the ejected yields of unstable isotopes, and we
estimate the rates of such events and the likelihood that COSI detects
their $\gamma$-ray lines.

\section{Models of i-process sources and their $\gamma$-ray line fluxes}

Viable \ipr\ sources of $\gamma$-ray emission must produce
\ipr\ species abundantly and eject them on a timescale comparable to
or shorter than the species' half-life, so that the nuclear
$\gamma$-ray lines are not trapped within the optically thick stellar
interior. Two such sites studied in detail are very late thermal
pulses (VLTPs) in low-mass post-AGB stars \cite{herwig2001b} and the
recurrent He-shell flashes in rapidly-accreting white dwarfs (RAWDs)
\cite{denissenkov2017}.

Sakurai's object erupted in 1994 and became a born-again giant within
$\sim 2$ years, with observed mass ejections \cite{hajduk:05}. In the born-again evolution scenario
\cite{herwig1999,herwig2001a} a low- or intermediate-mass AGB star
rapidly loses most of its H-rich envelope as a planetary nebula, and
the He shell in the cooling CO white dwarf may subsequently undergo a
VLTP with H ingestion, triggering \ipr\ nucleosynthesis
\cite{herwig2011} and expansion. H ingestion into the flash convection
can trigger a Global Oscillation of Shell H ingestion
\cite[GOSH,][]{herwig2014}, producing dynamic ejections of He-shell
material as observed in H-deficient knots around some late-thermal
pulse candidates
\cite{wesson2003,wesson2008,montoro-molina2023,fang2014}. Stellar
evolution computations, supported by observations, suggest that a VLTP
may occur in approximately 20\% of post-AGB objects
\cite{bertolami2024}.

In these VLTP scenarios the ejection is hydrodynamic rather than
quasi-static, and the expanding ejecta are expected to become
optically thin to the MeV $\gamma$-ray photons considered here on
timescales short compared to the relevant radioactive lifetimes (End
Matter).

RAWDs are considered the best candidates for the single-degenerate
channel leading to the explosion of Chandrasekhar-mass Type Ia
supernovae \cite[e.g.,][and references therein]{han2004}. A RAWD is a
CO white dwarf in a close binary system that accretes H-rich material
from a main-sequence, subgiant, or red giant branch (RGB) companion at
approximately $10^{-7}\,\Msunyr$, so that the
accreted H burns steadily on its surface. The accumulating He shell
eventually undergoes a thermal flash, triggering the \iprn, envelope
expansion, and ejection of material, similar to the born-again
evolution pathway, except that RAWDs may experience multiple He-shell
flashes over time.
In low-mass RAWDs ($M_\mathrm{WD}\approx 0.7\,\Msun$) with an
accreted material metallicity of $\mathrm{[Fe/H]}\geq -1.55$
(see Table~\ref{tab:isos} for the [A/B] notation definition)
the retention efficiency is below 10\%
\cite{denissenkov2017,denissenkov2019}, so more than 90\% of the
accreted material is ejected into the circumstellar medium after
undergoing \ipr\ nucleosynthesis.

Detailed 1D stellar evolution and 3D hydrodynamical simulations of H-ingestion 
scenarios show that the convective-reactive astrophysical fluid dynamics 
responsible for \ipr\ nucleosynthesis often lead to violent, non-spherical 
outbursts, ultimately resulting in a split of the He convective zone triggered 
by the GOSH instability \cite{herwig2014}, terminating the {\em i}-process, 
and launchiung the mass ejections \cite{denissenkov2017,stephens2021,choplin2021}.

We estimate upper limits for the ejected \ipr\ yields and compute the
resulting $\gamma$-ray photon flux of each unstable isotope from its
ejected mass, decay lifetime, and distance via Eq.~(\ref{eq:flux})
(End Matter).

To identify unstable, relatively long-lived isotopes in the
\ipr\ ejecta with $\gamma$-ray lines in the 0.5-2 MeV range
potentially detectable by COSI at up to 500 or 1000 pc, we examined
the $\gamma$-ray line table at
\url{https://atom.kaeri.re.kr/old/gamrays.html}, based on the
Evaluated Nuclear Structure Data File
(\url{https://www.nndc.bnl.gov}), extracted the abundances of relevant
isotopes from our models, and applied Eq. (\ref{eq:flux}).

\begin{figure}
  \centering
  \includegraphics[width=\columnwidth]{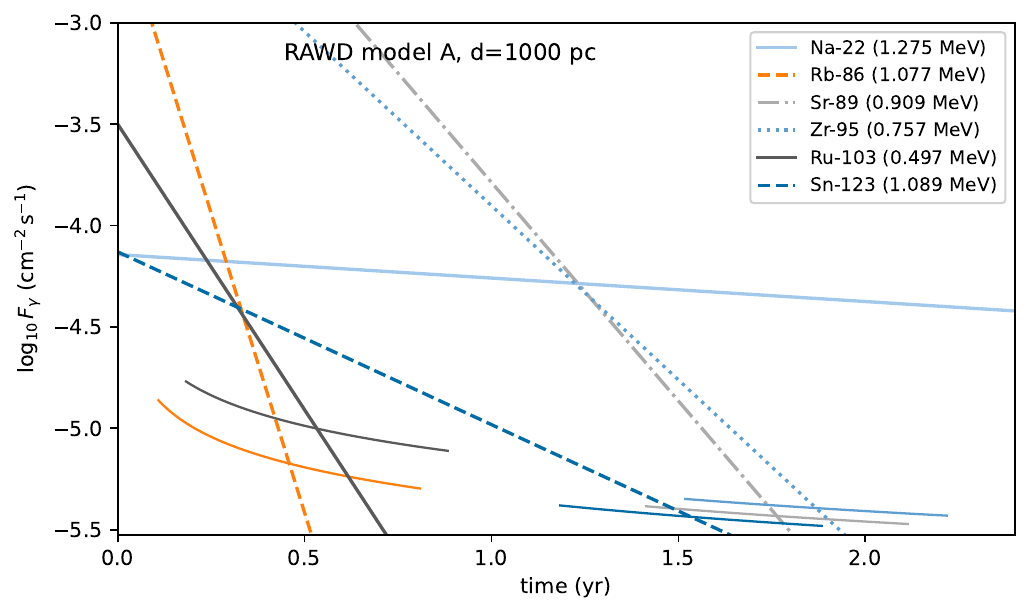}
  \caption{$\gamma$-ray line photon fluxes from the decay of unstable
    isotopes produced in the \iprn\ and ejected by our
    solar-metallicity RAWD model A at a distance of 1000 parsecs. The
    lower limit of the vertical axis corresponds to COSI's narrow-line
    sensitivity limit for energies between 0.5 and 2 MeV. Here
    ``sensitivity'' denotes the minimum detectable narrow-line flux at
    fixed detection significance, assuming background-limited
    performance and integration over the specified observation time.
    Short solid segments in each isotope's colour show the per-isotope
    COSI narrow-line sensitivity $S_{\rm iso}(t)=S(E_{\rm
    line},2\,{\rm yr})\sqrt{2\,{\rm yr}/t}$ near the crossover with
    $F(t)$, drawn over $\pm 0.35\,{\rm yr}$ of that crossover;
    integration up to $t_{\rm cross}$ accumulates enough signal for a
    $3\sigma$ COSI detection at the line energy. $^{22}$Na has no
    segment because $F(t)>S_{\rm iso}(t)$ throughout the mission and
    so remains detectable for its full duration. $S(E_{\rm
    line},2\,{\rm yr})$ values are taken from
    Fig.~\ref{fig:cosi_sensitivity_lines}. The $t^{-1/2}$
    extrapolation from the requirements-level 2-year curve is
    conservative at short integration times, where charged-particle
    activation buildup is incomplete. The rigorous detectability
    assessment using the time-averaged flux $\langle F\rangle_T$ and
    the optimal integration time $T^{\ast}_{\rm iso}$ is presented in
    Fig.~\ref{fig:cosi_sensitivity_lines}. The corresponding panels
    for the post-AGB VLTP model and for RAWD models B, C, and D, where
    in RAWD~D the long-lived $^{137}$Cs is sampled near its $\sim
    1\,{\rm yr}$ crossover, are provided in the Supplemental Material
    \cite{SuppMat}.}
  \label{fig:rawd_gamma_fluxes}
\end{figure}

Very few stars in the solar neighborhood have
$\mathrm{[Fe/H]} < -1.55$, so we focus on the post-AGB VLTP
model with $\mathrm{[Fe/H]} = -0.1$ \cite{denissenkov2018} and RAWD
models A, B, C, and D, which span a metallicity range from
$\mathrm{[Fe/H]} = 0$ to $-1.55$ (Table~\ref{tab:isos}) and exhibit
low mass retention efficiencies \cite{denissenkov2019}. We have
identified six unstable isotopes ejected by these models whose
$\gamma$-ray line fluxes could be detectable by COSI, assuming these
events occurred recently within 500, 1000, or 5000 pc from the Sun.
Figure~\ref{fig:rawd_gamma_fluxes} and the Supplemental Material
\cite{SuppMat} illustrate the magnitudes of these fluxes, which
decrease over time, together with the per-isotope COSI narrow-line
sensitivity $S_{\rm iso}(t)$ (see the caption of
Fig.~\ref{fig:rawd_gamma_fluxes}). The COSI sensitivity curve and
the detectability criterion based on time-averaged flux that is
adopted here are introduced at the start of Sec.~III
(Fig.~\ref{fig:cosi_sensitivity_lines}). The ejected masses of these
isotopes, along with their $\gamma$-line energies and lifetimes, are
listed in Table~\ref{tab:isos} (End Matter) for all RAWD models.

First, the post-AGB VLTP and RAWD A models produce
strong $^{22}$Na 1.275 MeV $\gamma$-line emissions, which could be
detectable by COSI over a period of several years. The potential
observation of this line from ONe novae has been discussed for decades
\cite{weiss1990, gehrz1998, jose1999, iyudin1999, starrfield2024},
but this line has not yet been detected. Our RAWD model A ejects more
$^{22}$Na than the highest-mass ONe nova model with the largest amount
of ejected $^{22}$Na from \cite{starrfield2024}.

Second, in all selected models both $^{89}$Sr and $^{95}$Zr produce
strong $\gamma$-line fluxes at 0.909 and 0.757 MeV, respectively,
detectable by COSI for up to two years after ejection.

Finally, in most of our selected models the $0.497$ MeV $^{103}$Ru and
$1.077$ MeV $^{86}$Rb lines, and in some the $1.089$ MeV $^{123}$Sn
line, can be detected by COSI over periods of $0.25$ to $1.5$ years.

The peak neutron densities during proton ingestion are comparable
across RAWD models A to D, but the initial metallicity determines the
neutron-to-seed ratio and thus the neutron exposure, the number of
neutrons captured per Fe-group seed nucleus. At lower metallicity, the
reduced seed abundance increases the exposure and shifts
nucleosynthesis from predominantly first-peak production toward
heavier nuclei. This behavior is reflected in the predicted yields,
which vary by up to nearly two orders of magnitude across the model
sequence (e.g., the ejected mass of $^{89}$Sr differs by almost a
factor of 40 between models A and D). Because different isotopes
respond differently to metallicity and neutron exposure, simultaneous
detection of multiple $\gamma$-ray lines (such as $^{22}$Na,
$^{89}$Sr, $^{95}$Zr, and $^{137}$Cs) would provide direct constraints
on the progenitor metallicity and the \iprn\ conditions, allowing a
sufficiently nearby event observed by COSI to discriminate between
competing RAWD models.

The predicted $^{137}$Cs production in RAWD model D reflects a larger
neutron exposure. In this regime, neutron captures proceed far from
the classical s-process path along the valley of stability but do not
reach the extreme conditions of the \rprn. A confirmed $\gamma$-ray
detection would therefore provide a direct, time-integrated probe of
nucleosynthesis in this intermediate neutron-density regime.

The post-AGB VLTP panel in the Supplemental Material \cite{SuppMat}
assumes a representative distance of 500 pc to illustrate the
detectability of a nearby future VLTP event. At the actual distance of Sakurai's object ($\approx
3$--$3.5\,\mathrm{kpc}$), the fluxes would be reduced by nearly a
factor of 50 and fall below COSI's sensitivity, and the decay of
short-lived isotopes such as $^{22}$Na since the 1994 eruption
suppresses any present-day $\gamma$-ray signal by several orders of
magnitude. Our results therefore represent predictions for a future
nearby VLTP event rather than the historical Sakurai outburst, whose
expected line fluxes were below the sensitivity of facilities such as
INTEGRAL/SPI, a non-detection consistent with instrumental limits
that motivates next-generation surveys.

\section{Likelihood of COSI detecting the predicted $\gamma$-ray lines}

\begin{figure}
  \centering
  \includegraphics[width=\columnwidth]{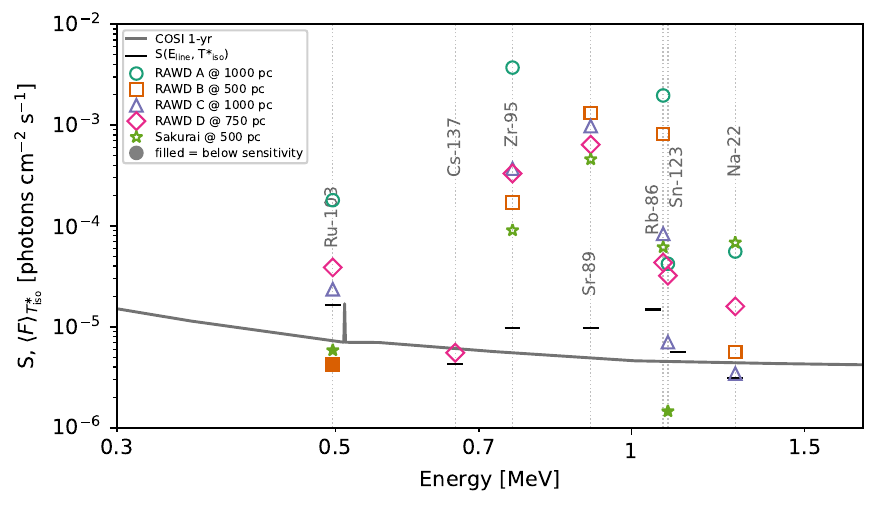}
  \caption{COSI 1-year, 3-$\sigma$, requirements-level narrow-line
    sensitivity (grey solid), obtained as $S(1\,{\rm yr}) =
    \sqrt{2}\,S(2\,{\rm yr})$ from the 24-month requirements curve
    (J. Tomsick, priv.\ comm., with the 511 keV spike at the Table~1
    requirement value of \cite{tomsick2024}), shown for reference.
    Vertical dotted lines mark the seven $\gamma$-ray line energies of the
    $i$-process isotopes considered here (Table~\ref{tab:isos}).  Short
    horizontal black ticks at each line energy show the species-specific
    sensitivity $S(E_{\rm line},T^{\ast}_{\rm iso})$ at the optimal
    background-limited integration time $T^{\ast}_{\rm iso} = 1.82\,t_{1/2}$
    (capped at 2~yr).  Coloured symbols show the corresponding
    time-averaged photon flux $\langle F\rangle_{T^{\ast}_{\rm iso}}$ at
    each line energy for every model at the reference distance adopted
    for that model in Fig.~\ref{fig:rawd_gamma_fluxes} (RAWD~A) or the
    Supplemental Material \cite{SuppMat} (post-AGB VLTP and
    RAWD~B--D).  A symbol above its corresponding black tick is detectable
    ($\langle F\rangle_{T^{\ast}_{\rm iso}}\geq S(E_{\rm line},T^{\ast}_{\rm
    iso})$; open marker); below the tick is not (filled marker).  The
    per-isotope ticks assume the event occurs at the start of the COSI
    mission. Events later in the mission have less integration time
    available, with sensitivity worse by approximately
    $\sqrt{T^{\ast}_{\rm iso}/T_{\rm avail}}$ plus an order-unity
    correction from charged-particle activation buildup.}
  \label{fig:cosi_sensitivity_lines}
\end{figure}

\begin{figure*}
  \centering
  \includegraphics[width=0.85\textwidth]{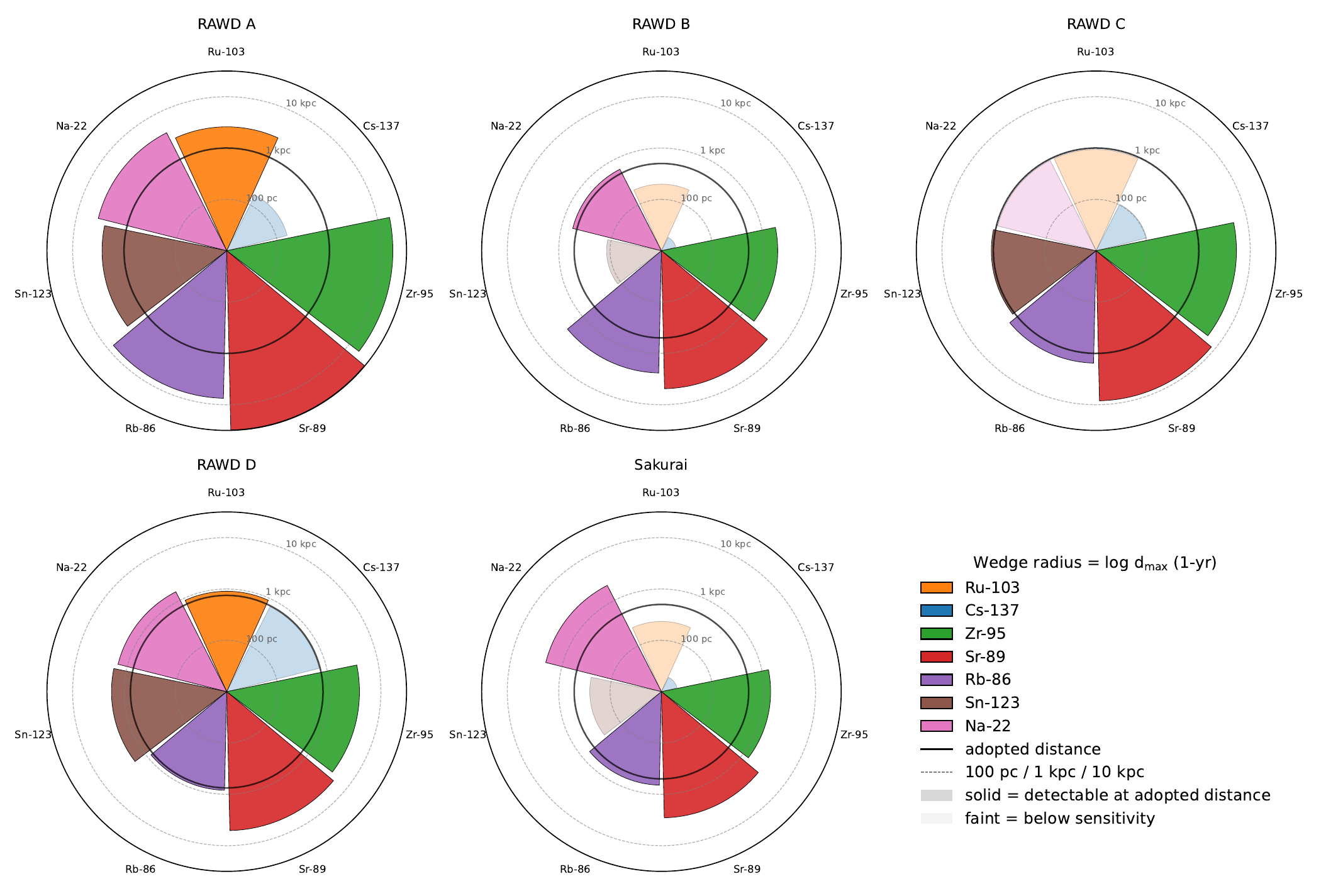}
  \caption{Maximum 1-year detection distance $d_{\max}$ for each
    $i$-process $\gamma$-ray line in every source model.  Each wedge's
    radius is $\log_{10}(d_{\max}/\mathrm{pc})$, with dashed concentric
    rings at 100~pc, 1~kpc, and 10~kpc.  The solid black ring labelled
    ``adopted distance'' is the distance adopted for that model in
    Fig.~\ref{fig:rawd_gamma_fluxes} (RAWD~A) or the Supplemental
    Material \cite{SuppMat} (Sakurai and RAWD~B--D).  Solid (faint) wedges
    indicate the line is (is not) detectable at that distance with a
    1-year integration.  Sensitivities are taken from the COSI curve of
    Fig.~\ref{fig:cosi_sensitivity_lines}.}
  \label{fig:detection_distance_pizza}
\end{figure*}

To assess the detectability of the predicted lines we compare the
time-averaged photon flux
$\langle F\rangle_T = (F_0\,\tau/T)\left[1 - \exp(-T/\tau)\right]$
over an integration window $T$ to the COSI narrow-line sensitivity
$S(E_{\rm line},T)$ at the line energy, with $T = 1\,$yr as the primary
reference window.  All seven $i$-process lines considered here fall in the
energy band where COSI is most sensitive
(Fig.~\ref{fig:cosi_sensitivity_lines}). Only the 0.497~MeV line of
$^{103}$Ru sits on the unfavourable shoulder of the 0.511~MeV
positron-annihilation feature.  The 1-yr sensitivity is obtained from the
2-yr COSI requirements curve of Ref.~\cite{tomsick2024} (Fig.~2a) by
background-limited scaling $S(T)\propto T^{-1/2}$.  This $\sqrt{t}$ scaling
is approximate, since charged-particle activation in the COSI detectors
builds up over the mission, but the deviation over the factor of $\sqrt{2}$
from 2 yr down to 1 yr is small.  At the adopted reference distances
(Fig.~\ref{fig:rawd_gamma_fluxes} and Supplemental Material
\cite{SuppMat}), the predicted $\langle F\rangle_{1\,\mathrm{yr}}$
exceeds the 1-yr sensitivity for most model/line pairs. The maximum 1-yr detection distance per
model/line pair is given in Fig.~\ref{fig:detection_distance_pizza}
(End Matter).

From the RAWD formation rate of binary population synthesis models and
the He-shell flash recurrence period, we estimate that the Milky Way
hosts approximately 1000 RAWDs, about 9 of them within 1 kpc of the
Sun (End Matter). The annual probability of a RAWD ejecting
\ipr\ products within 1 kpc is $\approx 0.018\%$, or $\approx 0.036\%$
over COSI's two-year prime mission. However, even at a distance of 5 kpc COSI could still detect
$\gamma$-ray lines from $^{86}$Rb, $^{89}$Sr, $^{95}$Zr, and
$^{103}$Ru recently ejected by a solar-metallicity RAWD (see
Fig.~\ref{fig:detection_distance_pizza}), with an annual probability
of $\approx 0.45\%$, corresponding to $\approx 0.9\%$ over the prime
mission. Furthermore, if the 0.909 MeV $^{89}$Sr line of RAWD model A
is detected within a few days of ejection, the maximum detection
distance increases to around 17.5 kpc even at COSI's reduced one-day
sensitivity, and 13--15 kpc are maintained for month-long or slightly
longer exposures, including a conservative 6-month transparency time
(End Matter). The annual detection probability then rises to
$\approx 5$--$5.5\%$, corresponding to $\approx 10$--$11\%$ over
COSI's two-year prime mission.

For Sakurai's object-like VLTP events, the three events recorded
within $\approx 5$ kpc over the past 100 yr imply an annual
probability of observing another similar event within 1 kpc of 0.12\%
(0.24\% over the prime mission), while an independent estimate from
the Galactic star formation rate gives 0.04\% (0.08\%) (End Matter).

These probabilities indicate that detecting $\gamma$-ray signals from
RAWDs is challenging but possible. Super-soft X-ray emission from
RAWDs is difficult to detect for the reasons discussed in Section 5 of
\cite{woods2016}, whereas $\gamma$-ray photons pass through
circumstellar material without significant attenuation, making them a
more reliable probe of \ipr\ nucleosynthesis in RAWDs.

For an extended mission beyond the two-year prime phase, the cumulative detection
probability would increase approximately linearly with observing time, given the low
event rates considered here.

\section{Conclusion}

We have identified six rare isotopes produced in \ipr\ nucleosynthesis
in our models of VLTPs in post-AGB stars and in RAWDs. Their
radioactive decays generate $\gamma$-ray lines that could be
detectable by COSI out to 5 kpc, provided such an event occurs during
its observational period. As emphasized in
recent reviews \cite{Wiedeking2025}, the i process represents an
emerging frontier in nuclear astrophysics, and MeV $\gamma$-ray
spectroscopy provides a uniquely direct test of this intermediate
neutron-density regime.

We excluded low-mass AGB stars \cite{choplin2022a} and post-RGB stars
\cite{battich2023} from our analysis because \ipr\ nucleosynthesis in
the former occurs only at $\mathrm{[Fe/H]}\leq -2.3$ and neither type
is expected to promptly eject \ipr\ products. However, second-peak \ipr\ patterns in
high-metallicity open clusters \cite{mishenina2015} suggest that
\ipr\ nucleosynthesis at nearly solar metallicity could produce
heavier \ipr\ elements, as proposed by \cite{karinkuzhi2023}.

Among these isotopes, $^{137}$Cs stands out
(Table~\ref{tab:isos}, final row). Its 0.662 MeV $\gamma$ line
produced in the RAWD D model results
in a photon flux that, averaged over 2 years, exceeds the 2-yr COSI
sensitivity at $0.662\,{\rm MeV}$ out to $d_{\max} \approx 850$~pc,
including the published 750~pc distance. On a 1-year integration this
same line is marginal at 750~pc ($d_{\max} \approx 720$~pc, see
Fig.~\ref{fig:detection_distance_pizza}). Because this line remains
detectable for roughly two decades following the \ipr\ event
\cite{SuppMat}, far exceeding COSI's two-year prime mission, the
cumulative detection probability in this low-rate regime scales with
the ratio of visibility time to mission duration, raising the
effective probability of observing a past RAWD D event within this
distance by more than an order of magnitude.

The probabilities of a post-AGB star undergoing a VLTP within 1 kpc or
a solar-metallicity RAWD experiencing a He-shell flash within 5 kpc
over COSI's two-year prime mission are both below 1\%. If such an
event does occur, the strongest and most long-lasting emissions are
expected from the radioactive decays of $^{22}$Na, $^{89}$Sr, and
$^{95}$Zr, at energies of $1.275$ MeV, $0.909$ MeV, and $0.757$ MeV,
respectively.

In the post-AGB VLTP and RAWD models, $^{22}$Na is produced by H
burning, similar to ONe nova models, not by neutron-capture reactions.
A simultaneous observation of $\gamma$-ray lines from decays of
$^{22}$Na and an \ipr\ rare isotope would therefore be direct evidence
of \iprn\ in H-ingestion He-shell flash events and the dynamic GOSH.
Given its relatively long half-life and large yields in our post-AGB
VLTP and RAWD A models, the 1.275 MeV line of $^{22}$Na could be
detected by COSI for about 10 years after the \ipr\ ends, increasing
its detection probability by a factor of 5.

The combination of short half-lives and rapid ejection makes our
identified $\gamma$-ray lines a characteristic signature of the
\iprn\ engine. The equilibrium abundances of short-lived nuclei such
as $^{89}$Sr, $^{95}$Zr, $^{103}$Ru, and $^{123}$Sn are negligible in
the classical s-process operating in low-mass AGB stars, where $\beta$
decay outpaces neutron capture at the modest neutron densities, and
even if produced, these isotopes would decay during the years-long
convective mixing and wind transport that brings He-shell material to
the interstellar medium. The post-AGB VLTP and RAWD scenarios studied
here are the only known stellar sites that combine \ipr\ neutron
densities \cite[the \ipr\ engine,][]{Wiedeking2025} with the prompt,
dynamically driven mass ejection of \cite[the GOSH,][]{herwig2014},
delivering the short-lived isotopes to the interstellar medium before
they decay. 

These neutron-rich species could in principle also be
\npr\ decay products from explosive He-shell burning in a
core-collapse supernova \cite{meyer2000}, but a nearby supernova would
be unambiguously identified by independent observations. A detection
of one of our $\gamma$-ray lines, particularly accompanied by the
$^{22}$Na line discussed above, would therefore constitute direct
evidence for an \ipr\ event with rapid ejection.

A non-detection over COSI's two-year prime mission carries diagnostic
weight only if a nearby post-AGB VLTP or RAWD He-shell flash
demonstrably occurred during the window. If such an event were
independently confirmed and no $\gamma$-ray lines were detected, the
constraint would bear on the astrophysics of the ejection (its
efficiency, timing, and degree of dynamical coupling to the underlying
nuclear engine) as much as on the predicted nuclear yields, which
already vary by 1--2 orders of magnitude across our model grid
(Table~\ref{tab:isos}).

Uncertainties of the underlying stellar models and of the nuclear
physics, including one-zone Monte Carlo estimates of the sensitivity
of our predicted yields to (n,$\gamma$) rate variations, are discussed
in the End Matter.

If future generations of $\gamma$-ray telescopes employing focusing
Laue or phased Fresnel lenses \cite{frontera2024,virgilli2022} achieve
a fiftyfold improvement in narrow-line sensitivity, the probability of
detecting at least some of these lines will increase to several tens
of percent. Liquid Ar detector concepts such as the GammaTPC
\cite{shutt2025}, with a mature design and promising fiducial
sensitivity, if confirmed, offer the additional advantage of all-sky
survey capability, capturing all detectable events without direct
pointing.

\begin{acknowledgments}
F.\ H.\ acknowledges funding through an NSERC Discovery Grant.
We thank Stefan Kimeswenger for input on the INTEGRAL non-detection of Sakurai's object, and John Tomsick for providing the underlying tabulated values of the COSI 24-month requirements-level sensitivity curve.
P.\ D.\ acknowledges CaNPAN support through NSERC under Grant
No. SAPPJ-797 2021-00032 ``Nuclear physics of the dynamic origin of
the elements''.
E.\ B.\ is supported by COSI, a NASA Small Explorer mission, under
NASA contract 80GSFC21C0059.
This work benefited from interactions and workshops co-organized by The Center for Nuclear astrophysics Across Messengers (CeNAM) which is supported by the U.S. Department of Energy, Office of Science, Office of Nuclear Physics, under Award Numbers DE-SC0023128 and DE-SC0026204.

\end{acknowledgments}

\section*{Data Availability}
The data and analysis code that support this Letter are
available on Zenodo, doi:\,\url{https://doi.org/10.5281/zenodo.20402090}.

\bibliography{paper_gammas}

\section*{End Matter}

\subsection*{Appendix A: $\gamma$-ray line flux calculation and ejecta transparency}

The convective-reactive instability of the VLTP and RAWD He-shell
flash scenarios rapidly transports \ipr\ products out of the optically
thick stellar interior. The H-deficient knots show that processed
intershell material reaches the circumstellar environment, and a
similar clumpy nova-ejecta morphology supports the picture of
radioactive nuclei residing in rapidly expanding, low-column-density
structures \cite{gomezgomar1998mnras}. The expanding ejecta are
therefore expected to become optically thin to MeV $\gamma$-ray
photons on timescales short compared to the relevant radioactive
lifetimes, with negligible subsequent circumstellar attenuation.

To estimate upper limits for \ipr\ yields, the masses of isotopes
ejected by the post-AGB VLTP and RAWD models into the circumstellar
medium, we multiply their undecayed mass fractions by the He-shell
masses, as derived from the models of \cite{denissenkov2018} and
\cite{denissenkov2019}, respectively. The mass fractions are averaged over the
He-shell convection zone and evaluated at the time when the maximum
abundance is reached, typically toward the end of the simulation.

The flux of $\gamma$-ray photons produced by the decay of an unstable
isotope ejected $t_\mathrm{y}$ years ago by an \ipr\ source at a
distance $d$ is given by
$$
F_{\gamma} = \frac{1}{4\pi d^2}\frac{M_\mathrm{ej}X}{AM_\mathrm{u}}
    \frac{\exp (-t_\mathrm{y}/\tau_\mathrm{y})}{t_\mathrm{sy}\tau_\mathrm{y}}\ (\mathrm{cm}^{-2}\mathrm{s}^{-1}),
$$
where $M_\mathrm{ej}$ is the ejected (He shell) mass, $X$, $A$ and
$\tau_\mathrm{y}$ are the mass-averaged abundance (mass fraction),
atomic mass, and lifetime in years ($\tau=t_{1/2}/\ln 2$) of the
isotope, respectively, $M_\mathrm{u}$ is the atomic mass unit,
$t_\mathrm{sy}$ is the number of seconds in one year. Substituting
numerical constants, the equation becomes
\begin{eqnarray}\label{eq:flux}
F_{\gamma}  & = \frac{0.01269}{(d_\mathrm{pc}/500\,\mathrm{pc})^2}\left(\frac{M_\mathrm{ej}
X}{10^{-8}\,\Msun}\right) \nonumber \\
    & \times\frac{\exp (-t_\mathrm{y}/\tau_\mathrm{y})}{A\tau_\mathrm{y}}\ (\mathrm{cm}^{-2}\mathrm{s}^{-1}),
\end{eqnarray}
where $d_\mathrm{pc}$ is the distance to the ejecta in parsecs (pc).

\subsection*{Appendix B: Occurrence rates}

During most of its lifetime, a rapidly-accreting white dwarf (RAWD)
steadily accretes H-rich material, which burns on its surface at high
temperature and luminosity, potentially producing super-soft X-ray
radiation. Several super-soft X-ray sources, possibly associated with
such systems, have been identified in the Large Magellanic Cloud
\cite{vandenheuvel1992,woods2016}. Binary population synthesis models
estimate the RAWD formation rate in the Milky Way galaxy to be on the
order of $10^{-3}\ \mathrm{yr}^{-1}$ \cite{han2004,benoit2018}. The
recurrence period for RAWD He-shell flashes is approximately $5\times
10^4$ years \cite{denissenkov2019}. Given their low retention
efficiencies, at least for $\mathrm{[Fe/H]}\geq -1.55$, RAWDs may
continue processing accreted H-rich material at a rate of $\sim
10^{-7}\,\Msunyr$ for several million
years. Therefore, we estimate that the Milky Way hosts approximately
1000 RAWDs, with about 9 of them located within 1 kpc of the Sun. This
estimate assumes that the spiral arms occupy 50\% of the Galactic disk
and that the disk has a radius of 15 kpc. Consequently, the annual
probability of a RAWD ejecting \ipr\ products within 1 kpc is $\approx
0.018\%$, and over COSI's two-year prime mission this corresponds to a
cumulative probability of $\approx 0.036\%$, assuming linear scaling
at low probabilities.

For the RAWD A model at 5 kpc and a typical line strength of $\sim
10^{-4}\ \mathrm{photons\,cm}^{-2}\mathrm{s}^{-1}$ (scaled from the
RAWD~A 1000~pc panel in Fig.~\ref{fig:rawd_gamma_fluxes}), the maximal
detection distance is given by $5\ \mathrm{kpc} \times
\sqrt{10^{-4}/S}$, where $S$ represents the sensitivity over a given
observation period. For a one-day observation, the sensitivity of
approximately $8 \times
10^{-5}\ \mathrm{photons\,cm}^{-2}\mathrm{s}^{-1}$ is comparable to
the signal strength, leaving the detection distance largely
unchanged. With a one-month exposure, the detection distance
extends to approximately 13 kpc. A slightly longer exposure allows for
a detection distance of about 15 kpc. Thus, even when accounting for a
conservative 6-month transparency time, it is reasonable to maintain
the 15 kpc estimate. This increases the annual detection probability
of a $\gamma$-line signal from the RAWD to $\sim 5\%$, i.e., $\sim
10\%$ over COSI's two-year prime mission.

It is challenging to estimate the occurrence rate of Sakurai's
object-like events. On the one hand, unlike the multiple He-shell
flashes in RAWDs, post-AGB stars may undergo only a single VLTP,
making such events less frequent. On the other hand, approximately
20\% of white dwarf remnants from low- and intermediate-mass AGB stars
are expected to experience VLTPs, suggesting a higher probability of
their occurrence. The interplay of these factors complicates an
accurate determination of the event rate. Three VLTP events have been
recorded within $\approx 5$ kpc over the past 100 yr: V605 Aql,
HuBi~1 (IRAS~17514$-$1555), and Sakurai's object
\cite{guerrero2018,bertolami2024}. Based on this empirical data, the
annual probability of observing another similar event with COSI at a
distance of up to 1 kpc is estimated to be 0.12\%, corresponding to
0.24\% over COSI's two-year prime mission (linear scaling). We can
also estimate this probability using a different approach. Taking a
Galactic star formation rate of $2\,\Msunyr$
\cite[e.g.,][]{elia2022}, the Salpeter initial mass function, and
assuming that stars with initial masses between $1\,\Msun$ and $8\,\Msun$
evolve into post-AGB white dwarfs, 20\% of which undergo a
VLTP, we estimate an annual probability for a VLTP event occurring
within 1 kpc of the Sun of 0.04\%, corresponding to 0.08\% over COSI's
two-year prime mission.

Optical counterparts of VLTPs and RAWD He-shell flashes, and known
born-again and late-thermal-pulse systems as targets for directed
$\gamma$-ray searches, are discussed in the Supplemental Material
\cite{SuppMat}.

\subsection*{Appendix C: Model uncertainties}

Stellar evolution models of RAWDs and post-AGB VLTP stars predicting
$\gamma$-ray line fluxes are still being developed and refined. As
discussed in \cite{Wiedeking2025} the convective-reactive nature of
the \ipr\ engine implies tight integration of nuclear energy
generation and turbulent advection, an inherently three-dimensional
problem that cannot be fully captured by 1D stellar evolution models.
The resulting uncertainties in the predicted \ipr\ yields are
difficult to quantify, but they could potentially impact the predicted
$\gamma$-ray line fluxes in both directions. For instance, if the GOSH
is less violent than predicted by current 3D hydrodynamical
simulations, the mass ejection and thus the \ipr\ yields could be
reduced. Conversely, if the GOSH is more violent, it could lead to
more efficient mixing and higher neutron densities, potentially
increasing the production of certain isotopes. There are also
additional potential sites of $\gamma$-ray emission associated with
\ipr\ that we have not considered here, such as super-AGB stars, for
which dynamic mass ejections have explicitely been speculated about
\cite{Jones2016}.

We performed one-zone Monte Carlo (MC) simulations at a constant
neutron density of $3.16\times 10^{14}\ \mathrm{cm}^{-3}$, where the
relevant (n,$\gamma$) reaction rates were randomly varied within
ranges defined by their default values from our \ipr\ nucleosynthesis
computations, divided and multiplied by maximum variation factors
$v_i^\mathrm{max}$, as described in \cite{denissenkov2021}. For the
unstable isotopes discussed here, the MC runs revealed
correlations only between the predicted abundances of
$^{95}$Zr and $^{89}$Sr and the variations of the
$^{95}$Y(n,$\gamma)^{96}$Y and $^{89}$Rb(n,$\gamma)^{90}$Rb reaction
rates, respectively. However, when adopting the
$v_i^\mathrm{max}$ values estimated in \cite{martinet2024}, or even using
our potentially overestimated values of $v_i^\mathrm{max}$ derived
with a method similar to that in \cite{denissenkov2018}, the
variations in the predicted abundances of $^{95}$Zr and $^{89}$Sr
remain below 30\%, implying that for practical reasons our predictions
are independent of nuclear physics uncertainties.

\begin{table*}
\caption{\label{tab:isos}
Yields of unstable isotopes $M_\mathrm{ej}X$ ($\Msun$) for different stellar \ipr\ nucleosynthesis models.}
\begin{ruledtabular}
\begin{tabular}{cccccccccc}
 Isotope & $E_\gamma\ (\mathrm{MeV})$ & $\tau_y\,\mathrm{(yr)}$ & post-AGB VLTP\footnotemark[1] &
 RAWD A\footnotemark[2] & RAWD B & RAWD C & RAWD D & RAWD E & RAWD G\\
\hline
 $^{22}$Na & $1.275$ & $3.75$ & $5.74\times 10^{-9}$ & $1.87\times 10^{-8}$ & $4.73\times 10^{-10}$ &
$1.15\times 10^{-9}$ & $3.01\times 10^{-9}$ & $1.70\times 10^{-9}$ & $4.10\times 10^{-9}$\\
 $^{86}$Rb & $1.077$ & $0.073$ & $5.39\times 10^{-10}$ & $6.90\times 10^{-8}$ & $7.13\times 10^{-9}$ &
$2.93\times 10^{-9}$ & $8.55\times 10^{-10}$ & $1.86\times 10^{-10}$ & $1.92\times 10^{-11}$\\
 $^{89}$Sr & $0.909$ & $0.20$ & $1.14\times 10^{-8}$ & $1.32\times 10^{-6}$ & $3.30\times 10^{-8}$ &
$9.73\times 10^{-8}$ & $3.58\times 10^{-8}$ & $9.03\times 10^{-9}$ & $1.61\times 10^{-9}$\\
 $^{95}$Zr & $0.757$ & $0.25$ & $3.02\times 10^{-9}$ & $4.96\times 10^{-7}$ & $5.71\times 10^{-9}$ &
$4.93\times 10^{-8}$ & $2.48\times 10^{-8}$ & $6.31\times 10^{-9}$ & $1.14\times 10^{-9}$\\
 $^{103}$Ru & $0.497$ & $0.15$ & $1.29\times 10^{-10}$ & $1.59\times 10^{-8}$ & $9.28\times 10^{-11}$ &
$2.08\times 10^{-9}$ & $1.92\times 10^{-9}$ & $4.92\times 10^{-10}$ & $9.35\times 10^{-11}$\\
 $^{123}$Sn & $1.089$ & $0.51$ & $1.26\times 10^{-10}$ & $1.47\times 10^{-8}$ & $2.80\times 10^{-11}$ &
$2.45\times 10^{-9}$ & $6.29\times 10^{-9}$ & $1.65\times 10^{-9}$ & $3.37\times 10^{-10}$\\
$^{137}$Cs & $0.662$ & $43.3$ & $4.85\times 10^{-11}$ & $3.05\times 10^{-9}$ & $4.36\times 10^{-11}$ &
$1.22\times 10^{-9}$ & $5.96\times 10^{-8}$ & $1.39\times 10^{-8}$ & $2.62\times 10^{-9}$\\
\end{tabular}
\end{ruledtabular}
\footnotetext[1]{From Ref.~\onlinecite{denissenkov2018}.}
\footnotetext[2]{RAWD models A, B, C, D, E, and G correspond to metallicities $\mathrm{[Fe/H]} =0, -0.7, -1.1, -1.55, -2.0$, and $-2.6$,
respectively, and are all from Ref.~\onlinecite{denissenkov2019}.
We use the standard stellar spectroscopy notation
[A/B]\,$=\log_{10}[N_\star(\mathrm{A})/N_\star(\mathrm{B})] -
\log_{10}[N_\odot (\mathrm{A})/N_\odot (\mathrm{B})]$, where $N_\star$
and $N_\odot$ are number densities of elements A and B in a star and
the Sun.}
\end{table*}

\clearpage
\setcounter{figure}{0}
\renewcommand{\thefigure}{S\arabic{figure}}
\setcounter{section}{0}
\renewcommand{\thesection}{S\Roman{section}}

\onecolumngrid
\begin{center}
  {\large\bfseries Supplemental Material:\\[2pt]
  Can COSI detect $\gamma$-ray lines from rare isotopes produced in
  the astrophysical intermediate neutron-capture process?}\\[6pt]
  Falk Herwig, Pavel Denissenkov, and Eric Burns
\end{center}
\twocolumngrid

\section{$\gamma$-ray line fluxes of the post-AGB VLTP model and RAWD models B, C, and D}

\begin{figure}
  \centering
  \includegraphics[width=\columnwidth]{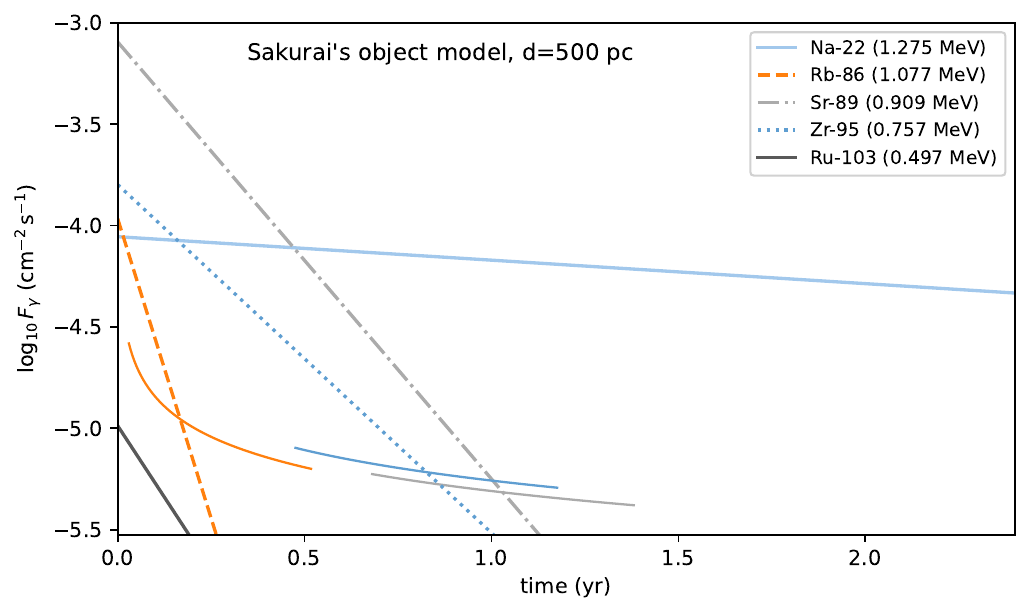}
  \caption{$\gamma$-ray line photon fluxes for the post-AGB VLTP model
    at a representative distance of 500 parsecs.}
  \label{fig:sakurai_sm}
\end{figure}

\begin{figure}
  \centering
  \includegraphics[width=\columnwidth]{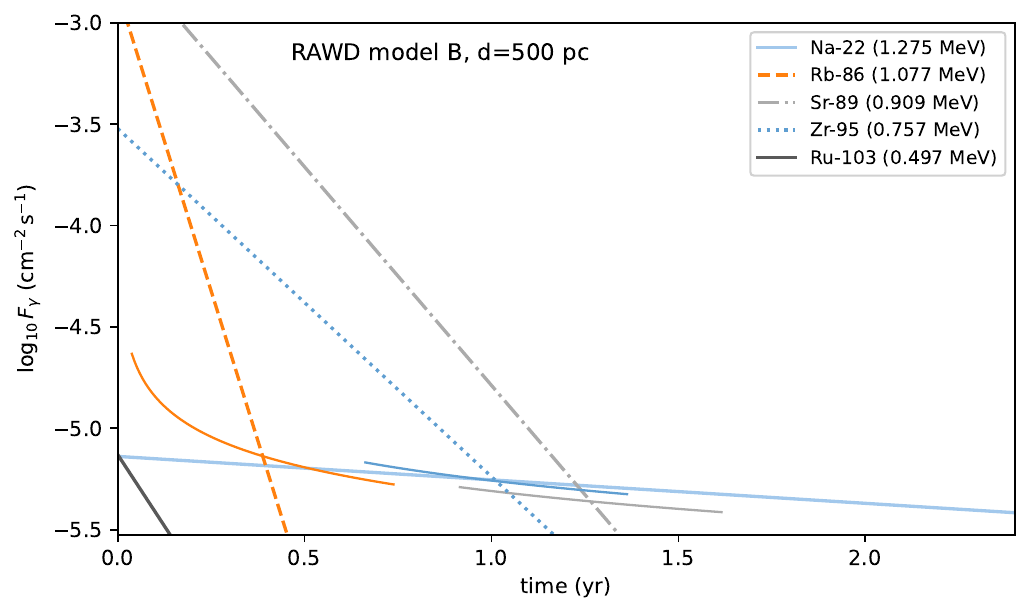}\\
  \includegraphics[width=\columnwidth]{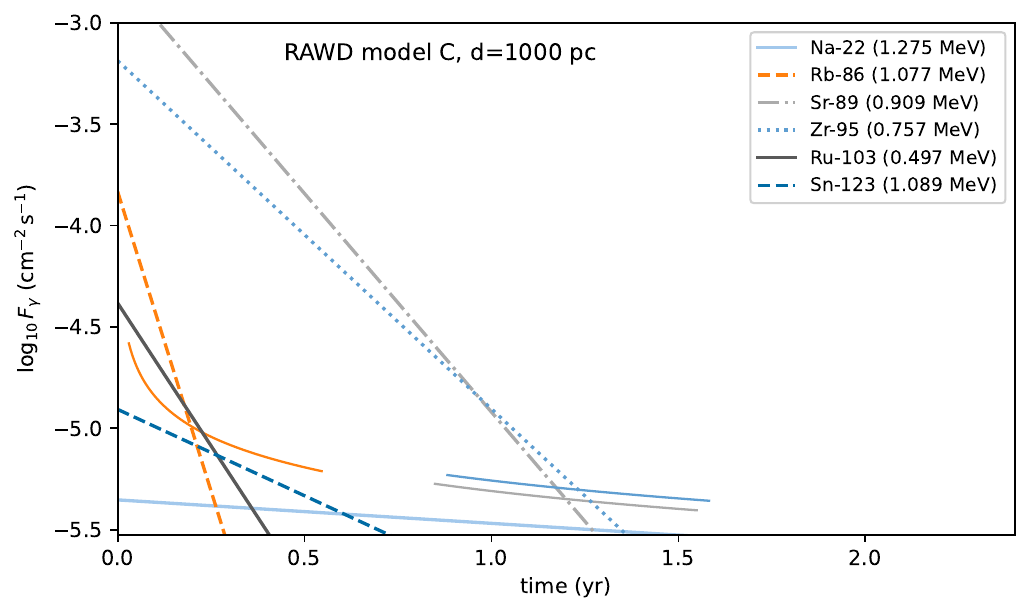}\\
  \includegraphics[width=\columnwidth]{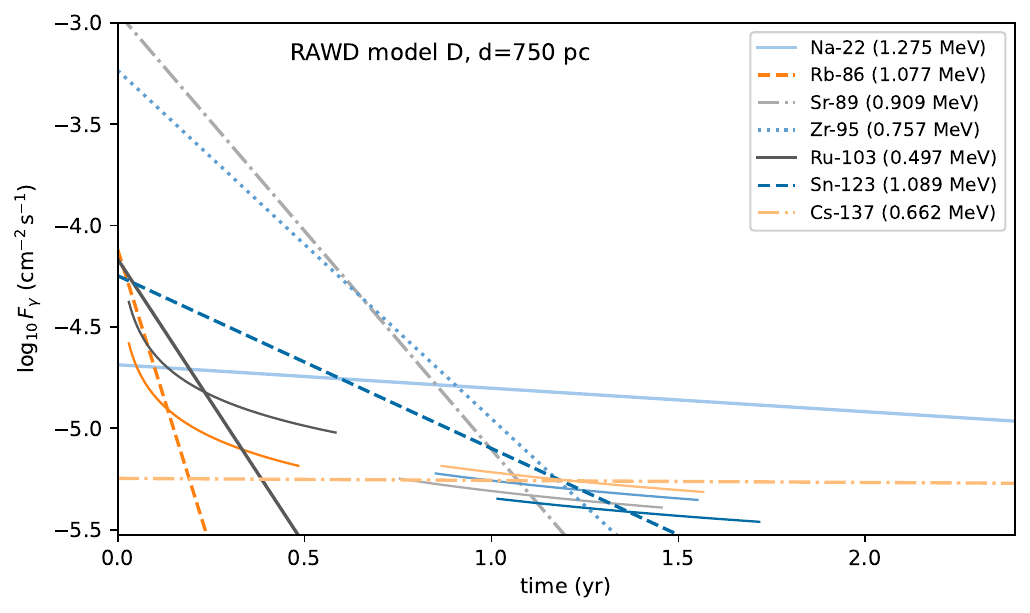}
  \caption{$\gamma$-ray line photon fluxes for RAWD model B at a
    distance of 500 parsecs (top), RAWD model C at 1000 parsecs
    (middle), and RAWD model D at 750 parsecs (bottom).}
  \label{fig:rawd_bcd}
\end{figure}

Figures~\ref{fig:sakurai_sm} and~\ref{fig:rawd_bcd} show the
$\gamma$-ray line photon fluxes from the decay of unstable isotopes
produced in the \iprn\ and ejected by our post-AGB VLTP model and by
RAWD models B, C, and D, complementing the RAWD model A panel shown in
Fig.~1 of the Letter. The post-AGB VLTP model assumes a representative
distance of 500 pc to illustrate the detectability of a nearby future
VLTP event. The adopted RAWD distances are 500 pc (model B), 1000 pc
(model C), and 750 pc (model D). The lower limit of the vertical axis
corresponds to COSI's narrow-line sensitivity limit for energies
between 0.5 and 2 MeV. As in Fig.~1 of the Letter, short solid
segments in each isotope's colour show the per-isotope COSI
narrow-line sensitivity $S_{\rm iso}(t)=S(E_{\rm line},2\,{\rm
yr})\sqrt{2\,{\rm yr}/t}$ near the crossover with $F(t)$, drawn over
$\pm 0.35\,{\rm yr}$ of that crossover. Integration up to $t_{\rm
cross}$ accumulates enough signal for a $3\sigma$ COSI detection at
the line energy. $^{22}$Na has no segment because $F(t)>S_{\rm
iso}(t)$ throughout the mission and so remains detectable for its full
duration. In RAWD model D the long-lived $^{137}$Cs is sampled near
its $\sim 1\,{\rm yr}$ crossover.

\section{X-ray and optical counterparts and targets for directed searches}

Section 5 of \cite{woods2016} discusses why super-soft X-ray sources
associated with RAWDs, particularly low-mass ones, are challenging to
detect. The key reasons include the high opacity of the accreted
material, self-absorption of X-rays, and potential obscuration by
circumstellar material. Variability in accretion rates and the
episodic nature of He-shell flashes can further complicate their
observability.

VLTPs and RAWD He-shell flashes are expected to have optical
counterparts. VLTP outbursts are discovered optically and reach peak
magnitudes of $m_V \sim 11$--15 at typical Galactic distances.
However, given the Galactic VLTP rates implied by the estimates in the
End Matter of the Letter, Rubin/LSST is expected to detect $\sim 0.3$
to 0.4 VLTPs over its 10-year survey, while RAWD flashes are even
rarer. Thus, optical triggering will not substantially increase event
statistics, but any detected VLTP would provide an immediate target of
opportunity for COSI.

Known born-again and late-thermal-pulse systems provide targets for
directed $\gamma$-ray searches. The most relevant recent VLTP object
is V4334~Sgr (Sakurai's object), whose 1996 outburst defines a
well-constrained epoch for decay-time modeling \cite{evans2020sakurai}.
HuBi~1 (IRAS~17514$-$1555) exhibits decades-long photometric evolution
consistent with a born-again pathway and remains a monitoring target
despite its larger distance \cite{guerrero2018}. SAO~244567 (the
Stingray Nebula) is generally interpreted as an LTP candidate
\cite{reindl2017stingray}. LTP events are a milder late He-shell flash
without clear evidence for strong H ingestion, and their
\ipr\ yields remain uncertain. In addition, older born-again planetary
nebulae such as A30 and A78 \cite{fang2014} and the historical VLTP
template V605~Aql \cite{clayton2013v605aql} serve as physical analogs
that constrain ejecta geometry and clump survival, although their
inferred event ages render them unsuitable for decade-lived
radionuclide searches.

\end{document}